\documentclass[12pt]{iopart}
\usepackage{iopams}  
\usepackage{amssymb}
\usepackage{bm}
\usepackage{epsfig}
\usepackage{wasysym}

\usepackage{color}

\begin{document}

\title{Edelstein and inverse Edelstein effects caused by the pristine surface states of topological insulators}

\author{Wei Chen}

\address{Department of Physics, PUC-Rio, Rio de Janeiro 22451-900, Brazil}

\ead{wchen@puc-rio.br}

\begin{abstract}

The Edelstein effect caused by the pristine surface states of three-dimensional topological insulators is investigated by means of a semiclassical approach. The combined effect of random impurity scattering and the spin-momentum locking of the gapless Dirac cone yields a current-induced surface spin accumulation independent from chemical potential and temperature. Through combing the semiclassical approach with the Bloch equation, the inverse Edelstein effect that converts the spin pumping spin current into a charge current is well explained. Consistency of these results with various experiments will be elaborated in detail.

\end{abstract}

%
%
%
%
%

\section{Introduction}

The peculiar electromagnetic response of three-dimensional (3D) topological insulators (TIs) point to their practical applications, especially to spintronic devices\cite{Pesin12_2}. Of particular interest are the spintronic effects caused by the surface states of 3D TIs, whose low energy sector is described by a two-dimensional (2D) Dirac Hamiltonian\cite{Fu07_2,Moore07,Roy09}. For instance, an inverse spin galvanic effect at the TI/ferromagnetic insulator interface due to the gapped surface state has been proposed\cite{Garate10}, as well as the spin Seebeck effect\cite{Okuma17} and the unique domain wall magnetization dynamics\cite{Tserkovnyak12,Kurebayashi19} in these system. Drawing analogy with other 2D parabolic band systems with strong Rashba spin-orbit coupling\cite{Silsbee01,Manchon08,Manchon09,Gambardella11}, the spin-momentum locking\cite{Zhang09,Liu10,Yazyev10} of the surface states is expected to yield a spin accumulation in the presence of an electric field or electric current, known as the Edelstein effect. Various experimental evidences indeed point to the existence of Edelstein effect at the surfaces of 3D TIs\cite{Tian15,Kondou16,Liu18,Dankert18}. Exploiting such effect as a mechanism for magnetization switching has been demonstrated in 3D TI and ferromagnetic metal (FMM) hybrid structures\cite{Mellnik14,Wang15,Mahendra18}. In the reciprocal process of what described above, a magnetization dynamics induced by ferromagnetic resonance in the FMM injects a spin current into a nearby TI, a phenomenon known as the spin pumping. Through the inverse Edelstein effect, the spin pumping spin current can be converted into a charge current, which has also been demonstrated and investigated intensively\cite{Shiomi14,Wang16_2,RojasSanchez16,Mendes17}. 


Despite the similarity with conventional 2D parabolic band Rashba systems, experiments reveal several puzzling features for the spin to charge interconversion mediated by the surface state. By varying the chemical potential $\mu$ through the Dirac cone, the efficiency of charge to spin current conversion is found to remain constant in a wide range of $\mu$, except a reduction near the Dirac point\cite{Kondou16}. Moreover, the effect is nearly constant of temperature from $0$ to $300$K\cite{Dankert18}. Secondly, the efficiency of the interconversion (see Sec.~\ref{sec:inverse_Edelstein} for the definitions) are of the order of nm\cite{Kondou16,RojasSanchez16,Zhang16_2}, which awaits a microscopic explanation. 



In this work, we elaborate that the surface disorder helps to understand a number of puzzling experimental results. Focusing on the situations in which the surface state Dirac cone remains gapless, we use the semiclassical Boltzmann equation and Bloch equation to give a microscopic account for the Edelstein and inverse Edelstein effects. The difference between the TI-based spintronics and the conventional metallic spintronics will be emphasized, and the consistency with various experimental results, using realistic material parameters\cite{Zhang09,Chen09}, will be demonstrated explicitly. 



This article is structured in the following manner. In Sec.~\ref{sec:Edelstein_STT}, we first incorporate the scattering of random surface impurities into a semiclassical approach to demonstrate a temperature and chemical potential independent Edelstein effect. Section \ref{sec:spin_pumping_inverse_Edelstein} first gives a microscopic mechanism for the spin pumping based on the Bloch equation, and then combine it with a semiclassical approach to demonstrate the inverse Edelstein effect. Section \ref{sec:conclusions} summarizes these results.

\section{Edelstein effect\label{sec:Edelstein_STT}}


We start by considering the pristine surface state in an isolated, semi-infinite 3D TI of symmetry class AII, described by the Hamiltonian and the spin-momentum locking\cite{Liu10,Shan10,Zhang12} 
\begin{eqnarray}
&&H=v_{F}k_{y}\sigma_{x}-v_{F}k_{x}\sigma_{y}={\bf d}\cdot{\boldsymbol\sigma}\;.
\nonumber \\
&&\langle{\boldsymbol\sigma}\rangle_{\bf k\pm}\equiv\langle \psi_{\bf k\pm}|{\boldsymbol\sigma}|\psi_{\bf k\pm}\rangle=\pm{\hat{\bf d}}=\pm{\hat{\bf x}}\sin\alpha\mp{\hat{\bf y}}\cos\alpha\;.
\label{e_h_spin_at_k}
\end{eqnarray} 
where $|\psi_{\bf k+}\rangle$ denotes eigenstate of the hole cone with energy $E_{\bf k+}=|{\bf d}|=v_{F}k$, whereas $|\psi_{\bf k-}\rangle$ is that of the electron cone of energy $E_{\bf k-}=-|{\bf d}|=-v_{F}k$. This two-component effective Hamiltonian is inherited from the full four-component low energy Hamiltonian in realistic TIs, as detailed in Appendix A. Given the density of states (DOS) $\rho(E)=a^{2}|E|/2\pi v_{F}^{2}$ and a random uncorrelated disorder $\langle\langle V({\bf r})V({\bf r}')\rangle\rangle=n_{i}V^{2}\delta({\bf r-r}')\equiv\gamma\delta({\bf r-r}')$, where $\langle\langle...\rangle\rangle$ denotes the impurity averaging, the mean free time\cite{Burkov10,Adroguer12} 
\begin{eqnarray}
&&\frac{1}{\tau(E)}=a^{2}\int\frac{d^{2}{\bf k}'}{(2\pi)^{2}}\frac{2\pi}{\hbar}\sum_{\beta}\langle\langle|\langle\psi_{\bf k'\beta}|V|\psi_{\bf k\beta}\rangle|^{2}\rangle\rangle\delta(E-E_{\bf k'\beta})
\nonumber \\
&&=\pi\gamma\rho(E)/\hbar\;.
\label{mean_free_time}
\end{eqnarray}
is inversely proportional to the DOS, a feature very different from other 2D parabolic band systems\cite{Manchon08,Manchon09}. In the presence of an electric field ${\boldsymbol{\mathcal E}}\parallel{\hat{\bf x}}$ and at finite chemical potential $\mu$, the semiclassical equation of motion of an electron in the hole cone $\beta=+$ and that in the electron cone $\beta=-$ modifies the Fermi distribution, which can be solved by incorporating the random impurity scattering into Boltzmann equation\cite{Adroguer12} 
\begin{eqnarray}
\frac{d{\bf k}}{dt}\cdot{\boldsymbol\nabla}_{\bf k}f_{\beta}&=&\frac{2\pi a^{2}}{\hbar}
\int\frac{d^{2}{\bf k}}{(2\pi)^{2}}\langle\langle|\langle\psi_{\bf k'\beta}|V|\psi_{\bf k\beta}\rangle|^{2}\rangle\rangle
\delta(E_{\bf k'\beta}-E_{\bf k\beta})\left[f_{\bf k'\beta}-f_{\bf k\beta}\right]\;.
\nonumber \\
\label{Edelstein_Boltzmann_eq}
\end{eqnarray} 
The equation may be solved by expanding the distribution function to leading order
\begin{eqnarray}
f_{\beta}=n_{F\beta}+\frac{\partial n_{F\beta}}{\partial E_{\bf k\beta}}\overline{f}(\alpha)\;,
\label{Edelstein_f_expansion}
\end{eqnarray}
where $n_{F\beta}=1/(e^{(E_{\bf k\beta}-\mu)/k_{B}T}+1)$. Putting Eq.~(\ref{Edelstein_f_expansion}) into Eq.~(\ref{Edelstein_Boltzmann_eq}), and using ${\boldsymbol\nabla}_{\bf k}E_{\bf k\beta}=\beta v_{F}\hat{\bf k}$ and the ansatz $\overline{f}(\alpha)\propto\cos\alpha$ yield the solution\cite{Adroguer12} 
\begin{eqnarray}
f_{\beta}=n_{F\beta}+\beta\frac{\partial n_{F\beta}}{\partial E_{\bf k\beta}}\frac{4e{\mathcal E}_{x}}{k\tilde{\gamma}}\cos\alpha\;.
\label{Edelstein_f_solution}
\end{eqnarray}
We have defined a dimensionless variable
\begin{eqnarray}
1/\tilde{\gamma}=(v_{F}/Va)^{2}/n_{i}\;,
\label{cleanliness_factor}
\end{eqnarray}
that measures the cleanliness of the surface, where $\tilde{\gamma}$ is essentially the percentage of disorder multiplied by the square of the disorder potential relative to the Dirac cone energy. The larger is $1/\tilde{\gamma}$, the cleaner is the surface.

\begin{figure}[ht]
\begin{center}
\includegraphics[clip=true,width=0.9\columnwidth]{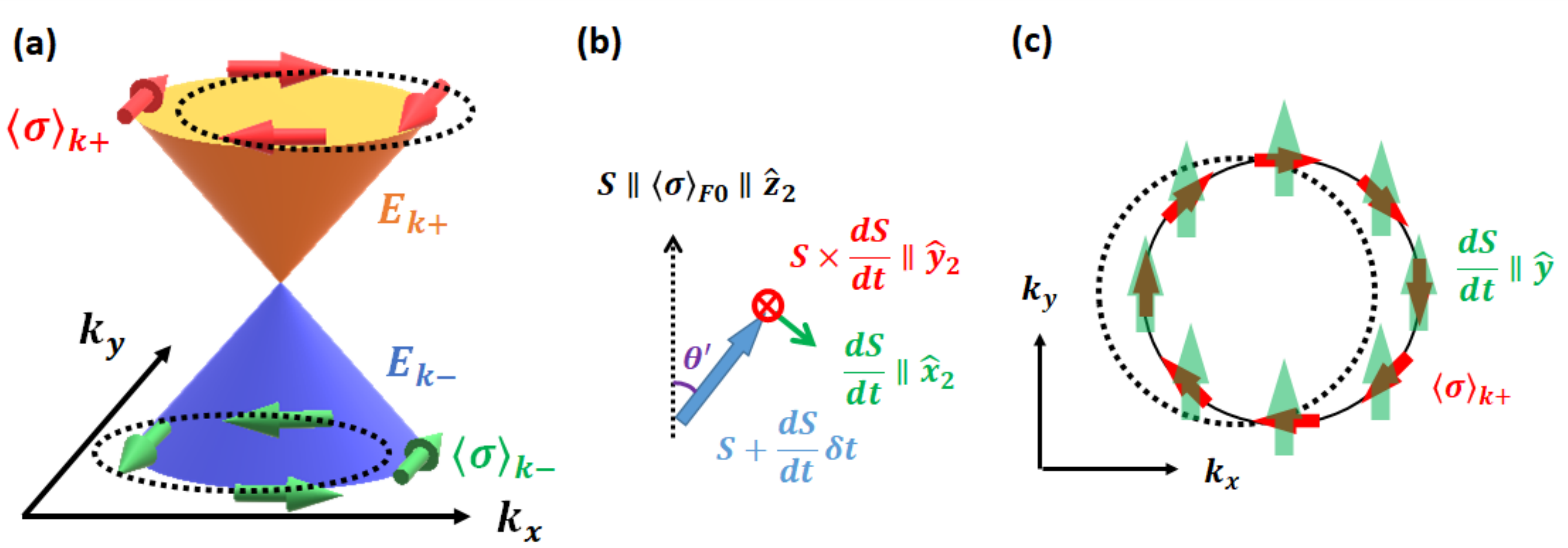}
\caption{(a) Schematics of the Edelstein effect. The electron cone and the hole cone of the surface state shift in opposite directions (dotted circles) in the presence of an in-plane electric field ${\boldsymbol{\mathcal E}}\parallel{\hat{\bf x}}$ due to their opposite group velocities, yet result in the same spin accumulation $\langle{\boldsymbol\sigma}\rangle\parallel{\hat{\bf y}}$ because of their opposite spin chirality (red and green arrows). (b) The rotated coordinate used in the spin pumping calculation. (c) Schematics of the inverse Edelstein effect caused by an instantaneous magnetization dynamics $d{\bf S}/dt\parallel{\hat{\bf y}}$ in the FMM. The generalized spin force $d{\bf S}/dt$ must project to the spin-momentum locked $\langle{\boldsymbol\sigma}\rangle_{\bf k\beta}$ in order to increase the population of $\langle{\boldsymbol\sigma}\rangle_{\bf k\beta}$, which yields a shift of the Fermi surface (dotted circle, assumed to be at the hole cone) and hence a charge current along ${\hat{\bf x}}$.} 
\label{fig:ISGE_schematics}
\end{center}
\end{figure}

Using Eq.~(\ref{e_h_spin_at_k}), the spin accumulation from each cone can be calculated from the Boltzmann equation 
\begin{eqnarray}
\langle{\boldsymbol\sigma}\rangle_{\beta}=a^{2}\int\frac{d^{2}{\bf k}}{(2\pi)^{2}}f_{\beta}\langle{\boldsymbol\sigma}\rangle_{\bf k\beta}
=\frac{a^{2}e{\mathcal E}_{x}}{\pi v_{F}\tilde{\gamma}}\left[\delta_{\beta=-}+\frac{\beta}{e^{-\mu/k_{B}T}+1}\right]{\hat{\bf y}}\;,
\label{spin_accumulation_hole_cone}
\end{eqnarray}
where the ${\hat{\bf x}}$ component vanishes due to the angular integration. In Eq.~(\ref{spin_accumulation_hole_cone}), the electron and the hole cone contribute the same sign of spin accumulation because, although the opposite group velocities ${\boldsymbol\nabla}_{\bf k}E_{\bf k\pm}=\pm v_{F}\hat{\bf k}$ cause the electron and hole cone to shift in opposite directions in respond to the external field, their spin expectation value along the shift is the same due to the opposite spin chiralities, as shown schematically in Fig.~\ref{fig:ISGE_schematics} (a) (one may as well define the shift without the group velocity part ${\boldsymbol\nabla}_{\bf k}E_{\bf k\pm}$, then for both cones $\beta=\pm$ it will be along the same direction\cite{Kondou16}). This surface spin accumulation occurs only when the external field ${\boldsymbol {\mathcal E}}$ has an in-plane component, and it is polarized along the direction perpendicular to the field, in contrast to the bulk magnetoelectric effect (ME) and the Edelstein effect at the TI/ferromagnetic insulator interface that give a magnetization along the field\cite{Qi08,Garate10}. The total spin accumulation $\sum_{\beta}\langle{\boldsymbol\sigma}\rangle_{\beta}=a^{2}e{\mathcal E}_{x}/\pi v_{F}\tilde{\gamma}$ is independent from temperature and chemical potential, and thus manifests even at zero temperature and zero chemical potential, in which case the diverging $\tau(E\rightarrow 0)$ compensates the vanishing Fermi surface. Note that the total spin accumulation $\sum_{\beta}\langle{\boldsymbol\sigma}\rangle_{\beta}$ is not determined by a single mean free time\cite{Zhang16_2} because the energy-dependent $\tau(E_{\bf k\beta})$ in Eq.~(\ref{mean_free_time}) has been integrated out in Eq.~(\ref{Edelstein_Boltzmann_eq}), leaving only the cleanliness factor $1/\tilde{\gamma}$. Our result well explains the temperature-independent surface spin accumulation recently measured in Bi$_{1.5}$Sb$_{0.5}$Te$_{1.7}$Se$_{1.3}$\cite{Dankert18}, and suggests that the Edelstein effect survives the surface band banding\cite{Bahramy12} and spatially modulated electron and hole puddles\cite{Beidenkopf11}, which may be simulated by a constant and a smoothly varying chemical potential, respectively.

The typical experimental current density $j_{c}\sim 10^{7}$A/cm$^{2}$ and conductivity $10^{7}$S/m correspond to the field strength ${\mathcal E}_{x}\sim 10^{4}$N/C. Assuming a lattice constant $a\sim$ nm and Fermi velocity $v_{F}\sim $ eVnm, the dimensionless spin polarization $a^{2}e{\mathcal E}_{x}/4\pi v_{F}\tilde{\gamma}\sim 10^{-6}/\tilde{\gamma}$ in units of $\hbar$ per unit cell is enhanced by the cleanliness factor in Eq.~(\ref{cleanliness_factor}). This conclusion is similar to that for the spin accumulation previously investigated in ultrathin TI films where the top and bottom surface states are coupled\cite{Siu17}. Given a reasonable value, say $\tilde{\gamma}\sim 0.01$, the spin polarization is few orders of magnitude larger that that induced by the bulk ME ${\bf M}\sim ea^{2}{\boldsymbol{\mathcal E}}/hc\sim 10^{-7}$\cite{Qi08}. Such a magnitude should be readily measurable by surface probes such as optical Kerr effect or X-ray magnetic circular dichroism, although existing experiments suggest that the surface magnetization may be highly interfered by the bulk bands contribution\cite{Liu18}. The effect is comparable to or may exceed the inverse spin galvanic effect at the TI/ferromagnetic insulator interface\cite{Garate10}, which in the same electric field strength yields an effective magnetic field ${\bf H}=\pm(1/2)\left(J_{pd}a/v_{F}S\right){\boldsymbol{\mathcal E}}e^{2}/h\;\;\Rightarrow{\bf B}\;\;\sim 0.1{\rm T}$ and subsequently a spin polarization ${\bf M}\sim\mu_{B}{\bf B}/\mu_{F}\sim 10^{-5}$, assuming the proximity induced exchange coupling that gaps out the Dirac cone is $J_{pd}\sim 0.1$eV. In comparison with the spin Hall effect in heavy metals, such as Pt and Ta with a spin Hall angle $\theta_{H}\sim 0.1$, the same current density and field strength produces a surface spin voltage $\mu_{\uparrow}-\mu_{\downarrow}\sim 0.01$meV and the corresponding surface spin polarization $(\mu_{\uparrow}-\mu_{\downarrow})/\mu_{F}\sim 10^{-5}$\cite{Zhang00,Chen16_quantum_tunneling}, which may also be surpassed by the Edelstein effect with sufficiently clean surface.

Finally, we remark that the charge current calculated out of this Boltzmann equation approach 
\begin{eqnarray}
j_{c}=-e\sum_{\beta}\langle v_{x}\rangle_{\beta}=-e\sum_{\beta}\langle\beta\frac{v_{F}}{\hbar}\cos\alpha\rangle=
\frac{{\mathcal E}_{x}}{\pi\tilde{\gamma}}\left(\frac{a^{2}e^{2}}{\hbar}\right)\;,
\label{Edelstein_charge_current}
\end{eqnarray}
is also determined by the surface cleanliness $1/\tilde{\gamma}$. Combining Eqs.~(\ref{spin_accumulation_hole_cone}) and (\ref{Edelstein_charge_current}), the spin accumulation is proportional to the charge current $\sum_{\beta}\langle{\boldsymbol\sigma}\rangle_{\beta}=\hat{\bf y}\hbar j_{c}/ev_{F}$ with a prefactor that only depends on the Fermi velocity but not the disorder. This behavior is consistent with the Edelstein effect predicted in the superconductor/TI/superconductor Josephson junction (also applicable to the normal state) using a more sophisticated Keldysh formalism\cite{Bobkova16}, which further supports our semiclassical approach.

\section{Spin pumping and inverse Edelstein effect \label{sec:spin_pumping_inverse_Edelstein}}

\subsection{Spin pumping \label{sec:spin_pumping}}

We now formulate the spin pumping caused by the surface state, which necessarily requires the consideration of a TI/ferromagnetic metal (TI/FMM) bilayer, since the spin pumping is experimentally induced by the ferromagnetic resonance (FMR) in the FMM layer. We will assume that the Dirac cone remains in the gapless pristine condition in the TI/FMM bilayers, so the formalism in Sec.~\ref{sec:Edelstein_STT} is applicable. The orbital degress of freedom of the FMM, which becomes an issue when it is made in proximity to the TI, is addressed in Appendix. Let us consider that the TI occupies the $z<0$ half-space, and the FMM film of finite thickness $0\leq z\leq\ell_{F}$. The principle is to solve the Bloch equation of the conduction electron spin in the FMM\cite{Zhang04,Chen15_STT}
\begin{eqnarray}
\frac{\partial\langle{\boldsymbol\sigma}\rangle_{F\beta}}{\partial t}+\partial_{z}{\boldsymbol j}_{\beta z}=\frac{J_{sd}}{\hbar}{\bf S}\times\langle{\boldsymbol\sigma}\rangle_{F\beta}-\overline{\boldsymbol\Gamma}_{\beta sf}\;,
\label{Bloch_equation}
\end{eqnarray}
in the presence of a magnetization ${\bf S}$ and magnetization dynamics $d{\bf S}/dt$, where $\overline{\boldsymbol\Gamma}_{\beta sf}$ is a spin relaxation term, $\langle{\boldsymbol\sigma}\rangle_{F\beta}$ is the conduction electron spin in the FMM contributed from $\beta$-cone, and ${\boldsymbol j}_{\beta z}$ is the spin current contributed from the $\beta$-cone at out-of-plane position $z$. To apply the Bloch equation to the dynamical magnetization, we assume that in equilibrium and at zero time $t=0$ where the magnetization is static, there is an equilibrium spin density parallel to the magnetization 
\begin{eqnarray}
\langle{\boldsymbol\sigma}(z,0)\rangle_{F\beta}=\langle{\boldsymbol\sigma}(z,0)\rangle_{F\beta 0}\parallel{\hat{\bf S}}\parallel{\hat{\bf z}}_{2}\;.
\end{eqnarray}
After the magnetization starts moving $d{\bf S}/dt\neq 0$, the spin density starts to deviate from its equilibrium value by 
\begin{eqnarray}
\langle{\boldsymbol\sigma}(z,\delta t)\rangle_{F\beta}=\langle{\boldsymbol\sigma}(z,0)\rangle_{F\beta 0}
+\delta\langle{\boldsymbol\sigma}(z,\delta t)\rangle_{F\beta}\;.
\end{eqnarray}
We use the rotated coordinate $(x_{2},y_{2},z_{2})$ that moves with the magnetization\cite{Chen15_STT} 
\begin{eqnarray}
{\hat{\bf x}}_{2}=\frac{1}{\left|\frac{d{\bf S}}{dt}\right|}\frac{d{\bf S}}{dt}\;,\;\;\;
{\hat{\bf y}}_{2}=\frac{1}{\left|\frac{d{\bf S}}{dt}\right|}{\hat{\bf S}}\times\frac{d{\bf S}}{dt}\;,
\end{eqnarray}
as shown in Fig.~\ref{fig:ISGE_schematics} (b). After the small time lapse $\delta t$, the angle that the magnetization makes with its equilibrium direction is 
\begin{eqnarray}
&&\lim_{\delta t\rightarrow 0}\sin\theta^{\prime}=\theta^{\prime}=\left|\frac{d{\hat{\bf S}}}{dt}\right|\delta t=\frac{1}{S}\left|\frac{d{\bf S}}{dt}\right|\delta t\;.
\end{eqnarray}
As long as the magnetization makes an angle $\theta^{\prime}$ with the equilibrium spin, the conduction electrons develop a spin deviation that is expected to be proportional to $\delta t$
\begin{eqnarray}
&&\delta\langle{\boldsymbol\sigma}\rangle_{F\beta}=\langle\sigma^{x_{2}}\rangle_{F\beta}{\hat{\bf x}}_{2}+\langle\sigma^{y_{2}}\rangle_{F\beta}{\hat{\bf y}}_{2}\propto\delta t\;.
\end{eqnarray}
The three terms in Eq.~(\ref{Bloch_equation}) then read
\begin{eqnarray}
&&\frac{\partial\langle{\boldsymbol\sigma}\rangle_{F\beta}}{\partial t}=\frac{\delta\langle{\boldsymbol\sigma}\rangle_{F\beta}}{\delta t}
=\frac{\langle\sigma^{x_{2}}\rangle_{F\beta}}{\sin\theta^{\prime}}\frac{d{\hat{\bf S}}}{dt}+\frac{\langle\sigma^{y_{2}}\rangle_{F\beta}}{\sin\theta^{\prime}}{\hat{\bf S}}\times\frac{d{\hat{\bf S}}}{dt}\;,
\nonumber \\
&&\frac{J_{sd}}{\hbar}{\bf S}\times\langle{\boldsymbol\sigma}\rangle_{F\beta}=\frac{\delta t}{\tau_{sd}}\left[-\frac{d{\hat{\bf S}}}{dt}\frac{\langle\sigma^{y_{2}}\rangle_{F\beta}}{\sin\theta^{\prime}}+{\bf S}\times\frac{d{\hat{\bf S}}}{dt}\frac{\langle\sigma^{x_{2}}\rangle_{F\beta}}{\sin\theta^{\prime}}\right]\;,
\nonumber \\
&&\overline{\boldsymbol\Gamma}_{\beta sf}=\frac{\delta\langle{\boldsymbol\sigma}\rangle_{F\beta}}{\tau_{sf}}=\frac{\delta t}{\tau_{sf}}\left[\frac{d{\hat{\bf S}}}{dt}\frac{\langle\sigma^{x_{2}}\rangle_{F\beta}}{\sin\theta^{\prime}}+{\hat{\bf S}}\times\frac{d{\hat{\bf S}}}{dt}\frac{\langle\sigma^{x_{2}}\rangle_{F\beta}}{\sin\theta^{\prime}}\right]\;,
\nonumber \\
\label{Bloch_eq_three_terms}
\end{eqnarray}
where $\tau_{sd}=\hbar/J_{sd}$ and $\tau_{sf}$ is the spin relaxation time. Since the last two terms in Eq.~(\ref{Bloch_eq_three_terms}) are proportional to $\delta t$ and vanish in the $\delta t\rightarrow 0$ limit, the Bloch equation reduces to the left hand side of Eq.~(\ref{Bloch_equation}), whose integration over the FMM slab yields the spin current flowing towards the TI
\begin{eqnarray}
a^{2}{\boldsymbol j}_{\beta 0}=a^{2}\int_{0}^{\ell_{F}}
dz\frac{\partial\langle{\boldsymbol\sigma}\rangle_{F\beta}}{\partial t}
=\frac{d\langle{\boldsymbol\sigma}\rangle_{\bf k\beta}}{dt}\;,
\label{spin_pumping}
\end{eqnarray}
where in the last line we have used angular momentum conservation to identify the spin current flowing into the TI with the time rate of the surface state spin. We see that the surface state spin expectation value $\langle{\boldsymbol\sigma}\rangle_{\bf k\beta}$ at momentum ${\bf k}$ changes because of the magnetization dynamics $d{\hat{\bf S}}/dt$, manifesting spin pumping. Moreover, at a time $t$ when the magnetization dynamics is a specific $d{\hat{\bf S}}/dt$, the time rate of such change is the same for all momenta ${\bf k}$ of the surface state that have the same energy $E_{\bf k\beta}$. 


\subsection{Inverse Edelstein effect based on spin pumping \label{sec:inverse_Edelstein}}

We proceed to discuss the charge current generated at the TI surface by the spin pumping in the TI/FMM bilayer, i.e., the inverse Edelstein effect based on spin pumping. Our starting point is that the time rate of the surface state spin, $d\langle{\boldsymbol\sigma}\rangle_{\bf k\beta}/dt$ in Eq.~(\ref{spin_pumping}), serves as a generalized spin force and yields the Boltzmann equation
\begin{eqnarray}
\frac{d\langle{\boldsymbol\sigma}\rangle_{\bf k\beta}}{dt}\cdot{\boldsymbol\nabla}_{\langle{\boldsymbol\sigma}\rangle}f_{\beta}&=&
\frac{2\pi a^{2}}{\hbar}
\int\frac{d^{2}{\bf k}}{(2\pi)^{2}}\langle\langle|\langle\psi_{\bf k'\beta}|V|\psi_{\bf k\beta}\rangle|^{2}\rangle\rangle
\nonumber \\
&&\times\delta(E_{\bf k'\beta}-E_{\bf k\beta})\left[f_{\bf k'\beta}-f_{\bf k\beta}\right]\;.
\label{spin_pumping_Boltzmann}
\end{eqnarray}
In essence, we treat $\langle{\boldsymbol\sigma}\rangle_{\bf k\beta}$ as part of the phase space of the Boltzmann equation. The derivative ${\boldsymbol\nabla}_{\langle{\boldsymbol\sigma}\rangle}$ is interpreted as the gradient along the direction that the population of $\langle{\boldsymbol\sigma}\rangle_{\bf k\beta}$ increases without changing the orientation of $\langle{\boldsymbol\sigma}\rangle_{\bf k\beta}$. It is given by
\begin{eqnarray}
{\boldsymbol\nabla}_{\langle{\boldsymbol\sigma}\rangle}f_{\beta}=
\langle{\boldsymbol\sigma}\rangle_{\bf k\beta}\frac{\beta}{a}\frac{\partial f_{\beta}}{\partial k}=\langle{\boldsymbol\sigma}\rangle_{\bf k}\frac{v_{F}}{a}\frac{\partial f_{\beta}}{\partial E_{\bf k\beta}}\;,
\end{eqnarray}
because first the generalized spin force $d\langle{\boldsymbol\sigma}\rangle_{\bf k\beta}/dt$ in Eq.~(\ref{spin_pumping_Boltzmann}) has to project to $\langle{\boldsymbol\sigma}\rangle_{\bf k\beta}$ to be accommodated by the state at ${\bf k}$, and because of the spin momentum locking, the increase of population of $\langle{\boldsymbol\sigma}\rangle_{\bf k\beta}$ is in the radial direction but opposite between the electron and the hole cone $\beta\partial/\partial k$, with $1/a$ inserted to keep the derivative dimensionless. This mechanism is shown schematically in Fig.~\ref{fig:ISGE_schematics} (c).


The same expansion and ansatz as Eq.~(\ref{Edelstein_f_expansion}) yields a average velocity along ${\hat{\bf x}}$ direction
\begin{eqnarray}
\sum_{\beta}\langle v_{x}\rangle_{\beta}=\sum_{\beta}a^{2}\int\frac{d^{2}{\bf k}}{(2\pi)^{2}}f_{\beta}\frac{v_{F}}{\hbar}\beta\cos\alpha\;,
\label{inverse_Edelstein_vx}
\end{eqnarray}
which is in general nonzero due to the nonequilibrium part of Eq.~(\ref{Edelstein_f_expansion}). Thus we see that a spin dynamics $d{\hat{\bf S}}/dt$ generates a charge current along ${\hat{\bf x}}$, manifesting spin pumping. We then argue the total spin current in the TI caused by spin pumping by noticing that the nonequilibrium part of Eq.~(\ref{Edelstein_f_expansion}) generates a spin accumulation 
\begin{eqnarray}
\sum_{\beta}\langle{\boldsymbol\sigma}\rangle_{\beta}=\sum_{\beta}a^{2}\int\frac{d^{2}{\bf k}}{(2\pi)^{2}}f_{\beta}\langle{\boldsymbol\sigma}\rangle_{\bf k\beta}\;,
\end{eqnarray}
In the presence of a phenomenological spin relaxation time $\tau_{sf}$, approximated as momentum independent, the total spin current is identifiable with $\sum_{\beta}\langle{\boldsymbol\sigma}\rangle_{\beta}/\tau_{sf}$. This total spin current and (\ref{inverse_Edelstein_vx}) allow to define the length scale that characterizes the efficiency of the spin current to charge current conversion in the inverse Edelstein effect\cite{RojasSanchez16,Zhang16_2} 
\begin{eqnarray}
\lambda_{IEE}=\left|\frac{\sum_{\beta}\langle v_{x}\rangle_{\beta}}{\sum_{\beta}\langle{\boldsymbol\sigma}\rangle_{\beta}/\tau_{sf}}\right|
=\frac{v_{F}\tau_{sf}}{\hbar}\;,
\end{eqnarray}
which takes the same form as a previous investigation\cite{Zhang16_2}, and is independent from the interface cleanliness, chemical potential, and temperature. To obtain the experimental value $\lambda_{IEE}\sim $nm requires $\tau_{sf}\sim 10$fs, which also agrees with the experimental value\cite{RojasSanchez16}.

\section{Conclusions \label{sec:conclusions}}

In summary, using a combination of semiclassical Boltzmann equation and Bloch equation, we give a theoretical account for several puzzling experimental results regarding the Edelstein and inverse Edelstein effects caused by the pristine surface state of 3D TIs. Owing to the scattering of random impurities, the energy-dependent mean free time, combined with the spin-momentum locking and linear DOS of the Dirac cone, render a current-induced surface spin accumulation that is independent from chemical potential and temperature\cite{Dankert18}. These results suggest that the Edelstein effect should survive the surface band banding\cite{Bahramy12} and spatial inhomogeneity\cite{Beidenkopf11}. For the spin to charge current conversion, i.e., inverse Edelstein effect based on spin pumping, the order of magnitude of the conversion efficiency $\lambda_{IEE}$ out of our semiclassical approach agrees with experiments\cite{RojasSanchez16}, and is found to be determined by the Fermi velocity and the spin relaxation time of the surface state. As our calculations clarify how the surface cleanliness and linear DOS affects the spin to charge interconversion mediated by the surface state, we anticipate that these results may help to engineer these effects for practical applications.

\section{Acknowledgement}

The author acknowledges fruitful discussions with M. H. Fischer and C. H. Lewenkopf. This project is partially supported by the incentives to research productivity fellowship in PUC-Rio.

\appendix

\section{Details of the low energy effective model for the surface state}

We now give an in-depth discussion about the surface state Hamiltonian according to the low energy effective theory of 3D TIs\cite{Zhang09,Liu10}. The low energy sector of realistic 3D TIs, such as Bi$_{2}$Se$_{3}$, are formed by the basis $|P1_{-}^{+},\uparrow\rangle$, $|P2_{+}^{-},\uparrow\rangle$, $|P1_{-}^{+},\downarrow\rangle$, $|P2_{+}^{-},\downarrow\rangle$, where the quantum numbers represent the hybridized Bi and Se orbitals, and the $\left\{\uparrow,\downarrow\right\}$ represents the spin index\cite{Liu10}. Using the representation for the $\Gamma$-matrices
\begin{eqnarray}
\Gamma_{i}=\left\{\sigma_{1}\otimes\tau_{1},\sigma_{2}\otimes\tau_{1},\sigma_{3}\otimes\tau_{1},
I_{\sigma}\otimes\tau_{2},I_{\sigma}\otimes\tau_{3}\right\},\;\;\;
\end{eqnarray} 
the low energy Hamiltonian obtained from ${\bf k\cdot p}$ theory is, keeping only the terms that are linear in momentum,
\begin{eqnarray}
\hat{H}=\tilde{H}_{0}+\tilde{H}_{1}\;,\;\;\;
\tilde{H}_{0}=M\Gamma_{5}+B_{0}\Gamma_{4}k_{z}\;,\;\;\;
\tilde{H}_{1}=A_{0}\left(\Gamma_{1}k_{y}-\Gamma_{2}k_{x}\right)\;,\;\;\;
\label{3D_TI_H0_H1}
\end{eqnarray}
where we have separated the planar $\tilde{H}_{1}$ and the out-of-plane component $\tilde{H}_{0}$. The surface state is solved by projecting the out-of-plane component into real space $\tilde{H}_{0}(k_{z}\rightarrow i\partial_{z})\psi(z)=E\psi(z)$. Because the out-of-plane component is block-diagonal 
\begin{eqnarray}
\tilde{H}_{0}=\left(
\begin{array}{cccc}
M & -iB_{0}k_{z} & & \\
iB_{0}k_{z} & -M & & \\
 & & M & -iB_{0}k_{z} \\
 & & iB_{0}k_{z} & -M
\end{array}
\right)\;,
\end{eqnarray}
the eigenvectors take the form 
\begin{eqnarray}
\Psi_{\uparrow}=\left(
\begin{array}{c}
\psi_{0} \\
{\bf 0}
\end{array}
\right)\;,\;\;\;
\Psi_{\uparrow}=\left(
\begin{array}{c}
{\bf 0} \\
\psi_{0}
\end{array}
\right)\;,
\label{3D_TI_surface_4_component}
\end{eqnarray}
with the two-component wave function satisfying 
\begin{eqnarray}
\left(M\tau_{3}-iB_{0}\tau_{2}\partial_{z}\right)\psi_{0}(z)=E\psi_{0}(z)\;. 
\label{3D_TI_H0_real_space}
\end{eqnarray}
Multiplying this equation by $\tau_{2}$, we see that the wave function is an eigenstate of $\tau_{1}\chi_{\eta}=\eta\chi_{\eta}=\pm\chi_{\eta}$. For the TI/FMM bilayer problem in Sec.~\ref{sec:spin_pumping_inverse_Edelstein} that considers the TI to occupy the $z<0$ half-space, we use the ansatz $\psi_{0}\propto e^{z/\xi}\chi_{\eta}$. Putting this ansatz into Eq.~(\ref{3D_TI_H0_real_space}) yields the solution for the zero energy $E=0$ surface state
\begin{eqnarray}
\xi=\eta\frac{B_{0}}{M}\;.
\end{eqnarray}
The eigenvalue $\eta=\pm$ is determined by the positiveness of the decay
length $\xi$ and the fact that only the topologically nontrivial 
phase has the surface state, which eventually is fixed by considering 
higher order expansions in the mass term\cite{Liu10}. Once $\eta$ is fixed, whether the spinor of the surface state is $\chi_{+}=(1,1)^{T}/\sqrt{2}$ or $\chi_{-}=(1,-1)^{T}/\sqrt{2}$ is also determined.

The effective Hamiltonian of the surface state in the $xy$-plane is given by considering the matrix elements in the basis of $\Psi=(\Psi_{\uparrow},\Psi_{\downarrow})$
\begin{eqnarray}
\langle\Psi|\Gamma_{1}|\Psi\rangle=\alpha\sigma_{x}\;,\;\;\;
\langle\Psi|\Gamma_{2}|\Psi\rangle=\alpha\sigma_{y}\;,
\end{eqnarray}
where $\alpha_{1}=\langle\psi_{0}|\tau_{1}|\psi_{0}\rangle$ is treated as a fitting parameter, such that the planar component in Eq.~(\ref{3D_TI_H0_H1}) has projected to the surface state yields
\begin{eqnarray}
\langle\Psi|\tilde{H}_{1}|\Psi\rangle
=\langle\Psi|A_{0}\left(\Gamma_{1}k_{y}-\Gamma_{2}k_{x}\right)|\Psi\rangle
=A_{0}\alpha_{1}\left(k_{y}\sigma_{x}-k_{x}\sigma_{y}\right)\;.
\label{3D_TI_surface_Heff}
\end{eqnarray} 
Identifying the Fermi velocity with $v_{F}=A_{0}\alpha_{1}$, we recover the surface state Hamiltonian in Eq.~(\ref{e_h_spin_at_k}).

The above discussion indicates that the basis of the surface state is the four-component eigenstate in Eq.~(\ref{3D_TI_surface_4_component}), and the Hamiltonian in Eq.~(\ref{3D_TI_surface_Heff}) acts on the spin space of this four-component basis. When the TI is made in contact with an FMM, as in the TI/FMM trilayer discussed in Sec.~\ref{sec:spin_pumping_inverse_Edelstein}, each spin species of the FMM conduction band is split into two sub-bands, owing to the hybridization with the $|P1_{-}^{+},\uparrow\rangle$, $|P2_{+}^{-},\uparrow\rangle$, $|P1_{-}^{+},\downarrow\rangle$, $|P2_{+}^{-},\downarrow\rangle$ orbitals of the TI. In other words, the proximity to the TI also makes the FMM to be described by a four-component basis. However, the Hamiltonian of the NM and FMM is degenerate for the two sub-bands, and hence can be expressed as 
\begin{eqnarray}
H_{N,F}=\left(
\begin{array}{cccc}
H_{\uparrow\uparrow} & & H_{\uparrow\downarrow} & \\
 & H_{\uparrow\uparrow} & & H_{\uparrow\downarrow} \\
H_{\downarrow\uparrow} & & H_{\downarrow\downarrow} & \\
 & H_{\downarrow\uparrow} & & H_{\downarrow\downarrow} 
\end{array}
\right)
=\left(\begin{array}{cc}
H_{\uparrow\uparrow} & H_{\uparrow\downarrow} \\
H_{\downarrow\uparrow} & H_{\downarrow\downarrow}
\end{array}
\right)\otimes I_{\tau}\;, 
\end{eqnarray}
where $I_{\tau}$ is the $2\times 2$ identity matrix in the orbital space, and the $H_{\sigma\sigma^{\prime}}$ acts on the spin space. The invariance of the Hamiltonian under exchange of the orbital indices $\left[H_{N,F},\tau_{1}\right]=0$ is very convenient, because it implies we can rotate the orbital wave function in the FMM to the basis that is an eigenstate of $\tau_{1}\chi_{\eta}=\eta\chi_{\eta}=\pm\chi_{\pm}$, same basis as the surface state wave function in the TI discussed after Eq.~(\ref{3D_TI_H0_real_space}), with the same eigenvalue $\eta$. In this basis, the orbital degrees of freedom does not explicitly manifest in the spintronic properties discussed in the present work, and hence we can ignore it for simplicity and only focus on the spin degrees of freedom. In other words, we may reduce the four-component basis in Eq.~(\ref{3D_TI_surface_4_component}) to a two-component one
\begin{eqnarray}
\Psi_{\uparrow}=\left(
\begin{array}{c}
\psi_{0} \\
{\bf 0}
\end{array}
\right)\rightarrow\left(
\begin{array}{c}
1 \\
0
\end{array}
\right),\;\;\;
\Psi_{\downarrow}=\left(
\begin{array}{c}
{\bf 0} \\
\psi_{0}
\end{array}
\right)\rightarrow\left(
\begin{array}{c}
0 \\
1
\end{array}
\right),\;\;\;\;\;\;\;\;
\label{3D_TI_surface_2_component}
\end{eqnarray}
to simplify the problem, and so follows the two-component formalism used throughout the article.

\vspace{1cm}

\bibliographystyle{unsrt}
\bibliography{Literatur}

\end{document}